\documentclass[aps,pra,preprint,superscriptaddress]{revtex4-1}

\usepackage{amsmath,amssymb}
\usepackage{bm}
\usepackage{color}
\usepackage{graphicx}

\begin{document}

\title{\large \bf  Magnetic-field-dependent angular distributions and linear polarizations of emissions from the $2p^53s~^3P_2$ state in Ne-like ions}

\author{Jiguang Li}
\altaffiliation{Present address: Data Center for High Energy Density Physics, Institute of Applied Physics and Computational Mathematics, PO Box 8009, Beijing 100088, China}
\email{Li\_Jiguang@iapcm.ac.cn}
\affiliation{Department of Physics, Lund University, 22100 Lund, Sweden}
\author{Tomas Brage}
\affiliation{Department of Physics, Lund University, 22100 Lund, Sweden}
\author{Per J\"onsson}
\affiliation{Materials Science and Applied Mathematics, Malm\"o University, 21120 Malm\"o, Sweden}
\author{Yang Yang}
\affiliation{The Key Laboratory of Applied Ion Beam Physics, Minister of Education, Shanghai 200433, China}
\affiliation{Shanghai EBIT Laboratory, Institute of Modern Physics, Fudan University, Shanghai 200433, China}

\date{\today}

\begin{abstract}
We studied the effect of an external magnetic field on the angular distributions and the linear polarizations of emissions from the $2s^22p^53s~^3P^o_2$ state in Ne-like ions. Since a B-dependent E1 decay channel is opened by the magnetic field and competes with the inherent M2 transition, the angular distributions and the linear polarizations strongly depend on the magnetic field strength (B). As an example, we illustrated the effect in Ne-like Mg. The B-dependent angular distributions and linear polarization degrees can also be considered as a tool for diagnostics of the magnetic field in plasmas.
\end{abstract}

\maketitle

Magnetic-field induced transitions (MITs), arising from mixing between atomic state wave functions (ASFs) with different total angular momentum $J$ due to an external magnetic field~\cite{Andrew1967, Wood1968}, have attracted attentions again, since which are considered as a tool to diagnose the strength of the magnetic field in plasmas~\cite{Beiersdorfer2003}. Lately, we investigated in detail the influences of MITs on the lifetimes of the $^3P^o_2$ and $^3P^o_0$ metastable states of Ne-like ions~\cite{Li2013}. It is worth noting that in this case there are two decay channels from the $^3P^o_2$ level to the ground state, that is, the magnetic quadrupole (M2) and magnetic-field induced electric dipole (E1) transitions, which give rise to the interference. However, this interference effect, although sensitive to the transition amplitudes involved and thus the strength of the magnetic field, can not be seen by lifetimes. In order to obtain more information about the influence of an external magnetic field on the radiative emission from the $^3P_2$ level, we furthermore studied the angular distributions and the linear polarizations of corresponding lines in Ne-like ions.

In general, the spontaneous emission probability from an initial state $\Psi_i$ to a final state $\Psi_f$ is given in the atomic unit by
\begin{equation}
\label{TP}
A_{if} = \frac{\alpha}{2 \pi} \omega_{ij} |T_{if}|^2, \\
\end{equation}
where $\alpha$ is the fine-structure constant and $\omega_{ij}$ is the frequency of the photon. According to the multipole expansion of the vector potential, the transition amplitude $T_{if}$ is written as~\cite{Johnson2007}
\begin{equation}
\label{TA}
T_{if} = 4\pi \sum_{JM\lambda}i^{J-\lambda}[{\bm Y}^{(\lambda)*}_{JM}(\vec{k}) \cdot \hat{\epsilon}][T^{(\lambda)}_{JM}]_{if},
\end{equation}
where
\begin{equation}
[T^{(\lambda)}_{JM}]_{if} = \int d^3 r \Psi^{\dagger}_i(\vec{r}) {\bm \alpha} \cdot {\bm a}^{(\lambda)}_{JM}(\vec{r})\Psi_f(\vec{r}).
\end{equation}
Here, ${\bm Y}^{(\lambda)}_{JM}(\vec{k})$ is the vector spherical harmonics~\cite{Varshalovich1998}. $\vec{k}$ stands for the direction of the photon's propagation in the coordinate space ($\vec{r}$) where the atom is located. The vector $\hat{\epsilon}$ describes the photon's polarization. ${\bm a}^{(\lambda)}_{JM}(\vec{r})$ is multipole potentials with $\lambda=0$ for the magnetic and $\lambda=1$ for the electric multipoles. ${\bm \alpha}$ is the vector of the Dirac matrices. The subscript $J$ represents the total angular momentum of the photon, and $M$ its $z$-component. It should be emphasized that all information about the photon's propagation direction and its polarization is accounted for by the expansion coefficients $[{\bm Y}^{(\lambda)*}_{JM}(\vec{k}) \cdot \hat{\epsilon}]$ in Eq. (\ref{TA}).
   
In the presence of an external magnetic field, it is convenient to choose the magnetic-field direction as the quantization axis. As a result, atomic state functions with the same parity and the magnetic quantum number $M$ but different total angular momentums $J$ are mixed, namely,
\begin{equation}
\label{ASFM}
| \Psi(M) \rangle = \sum_{\Gamma J} d_{\Gamma J} | \Phi(\Gamma J M) \rangle.
\end{equation} 
Here, \{$d_{\Gamma J}$\} denote mixing coefficients. In the first-order perturbation approximation, they can be expressed as 
\begin{equation}
d_{\Gamma J} = \frac{\langle \Phi(\Gamma J M) | H_m | \Phi(\Gamma_0 J_0 M_0) \rangle }{E(\Phi(\Gamma_0 J_0)) - E(\Phi(\Gamma J))},
\end{equation}
where the subscript $0$ denotes the state under investigation. The interaction Hamiltonian between the external magnetic field and a N-electron atom is written by~\cite{Andersson2008}
\begin{equation}
H_m = ({\bf N}^{(1)} + \Delta {\bf N}^{(1)}) \cdot {\bf B}
\end{equation}  
with
\begin{align}
{\bf N}^{(1)} &= \sum_q -i \frac{\sqrt{2}}{2\alpha} r_q ({\bm \alpha}_q{\bf C}^{(1)}(q))^{(1)}, \\
\Delta {\bf N}^{(1)} &= \sum_q \frac{g_s - 2}{2} \beta_q {\bf \Sigma}_q.
\end{align}
Here, the last term is the so-called Schwinger QED correction. ${\bm \alpha}$ and $\beta$ constitute the Dirac matrices, and ${\bf \Sigma} = \bigl( \begin{smallmatrix} {\bf \sigma} & 0 \\
                                                                               0 &  {\bf \sigma} \end{smallmatrix} \bigr)$ is the relativistic spin-matrix. The electron $g$ factor with the QED correction is $g_s=2.000232$. As can be seen from Eq. (\ref{ASFM}), new decay channels may be opened by the external magnetic field due to the mix of atomic state wave functions with different total angular momentum $J$. These lines are called magnetic-field induced transitions (MIT)~\cite{Li2013}.
 
For the case of Ne-like ions under consideration, we approximately express the initial (i) and final (f) atomic state involved, respectively, as~\cite{Li2013}
\begin{equation}
| \Psi(2p^53s~^3P_2 ~ M) \rangle_i  = d_0 | \Phi(2p^53s~^3P_2 ~ M) \rangle + d_1 | \Phi(2p^53s~^3P_1 ~ M) \rangle + d_2 | \Phi(2p^53s~^1P_1 ~ M) \rangle, 
\end{equation}
and
\begin{equation}
| \Psi(2p^6~^1S_0 ~ M) \rangle_f = | \Phi(2p^6~^1S_0 ~ M) \rangle.
\end{equation}
As can be seen from equations above, the external magnetic field gives rise to the mix between the $2p^53s~^3P_2$ state and the $2p^53s~^3P_1$ and $2p^53s~^1P_1$ states. However, it is worth nothing that the mix arises only for the levels with magnetic quantum number $M=0, \pm1$. As a result, there are two decay channels, that is, the magnetic-field induced E1 transition and the magnetic quadrupole (M2) transition, for $M=0, \pm 1$ magnetic sublevels. Moreover, these two decay channels can cause the interference effect. For other two sublevels belonging to $2p^53s~^3P_2$ only the M2 transition occurs. 
Regardless the photon's polarization, the transition rates from different magnetic sublevels in the $2p^53s~^3P_2$ state to the ground state are written as
\begin{scriptsize}
\begin{eqnarray}
\label{ADTP}
A_{if} 
= \frac{k}{2 \pi} \left\{
\begin{array}{l}
 \left|{\bm Y}^{(0)}_{2M}(\vec{k})\right|^2 \left|[T^{(0)}_{2M}]_{if} \right|^2 \ \ \ \mbox{for}\  M=\pm 2 \\
 \left|{\bm Y}^{(0)}_{2M}(\vec{k})\right|^2 \left|[T^{(0)}_{2M}]_{if}\right|^2 + \left|{\bm Y}^{(1)}_{1M}(\vec{k})\right|^2 \left|[T^{(1)}_{1M}]_{if}\right|^2 - 2[{\bm Y}^{(0)*}_{2M}(\vec{k}) \cdot {\bm Y}^{(1)}_{1M}(\vec{k})][T^{(0)}_{2M}]^*_{if}[T^{(1)}_{1M}]_{if}  \ \ \ \mbox{for}\ M=\pm 1 \\
\left|{\bm Y}^{(0)}_{2M}(\vec{k})\right|^2 \left|[T^{(0)}_{2M}]_{if}\right|^2 + \left|{\bm Y}^{(1)}_{1M}(\vec{k})\right|^2 \left|[T^{(1)}_{1M}]_{if}\right|^2  \ \ \ \mbox{for}\ M=0.
\end{array}
\right.
\end{eqnarray}
\end{scriptsize}
Note that the interference effect between the magnetic-field induced E1 and M2 decay channels only appears in the transitions from $M=\pm 1$, when the photon's polarization is neglected. In addition, one should keep in mind that the magnetic-field induced E1 transition rate is proportional to ${\rm B}^2$~\cite{Li2013}. This leads to the B-dependent angular distributions for $A_{if}$.
  
\begin{figure}[!th]
\includegraphics[scale=0.6]{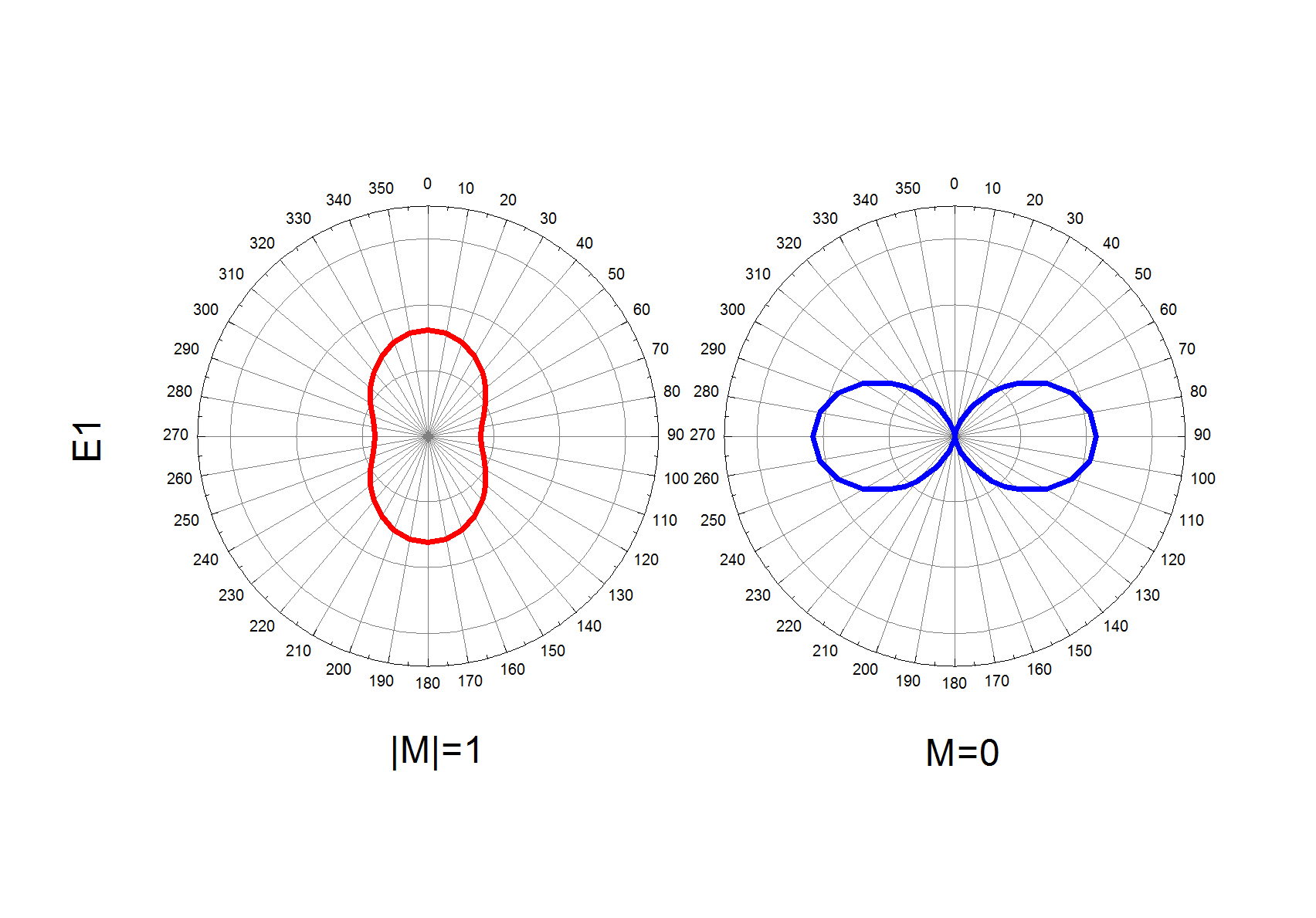}
\vspace{-2.5cm}
\includegraphics[scale=1.0]{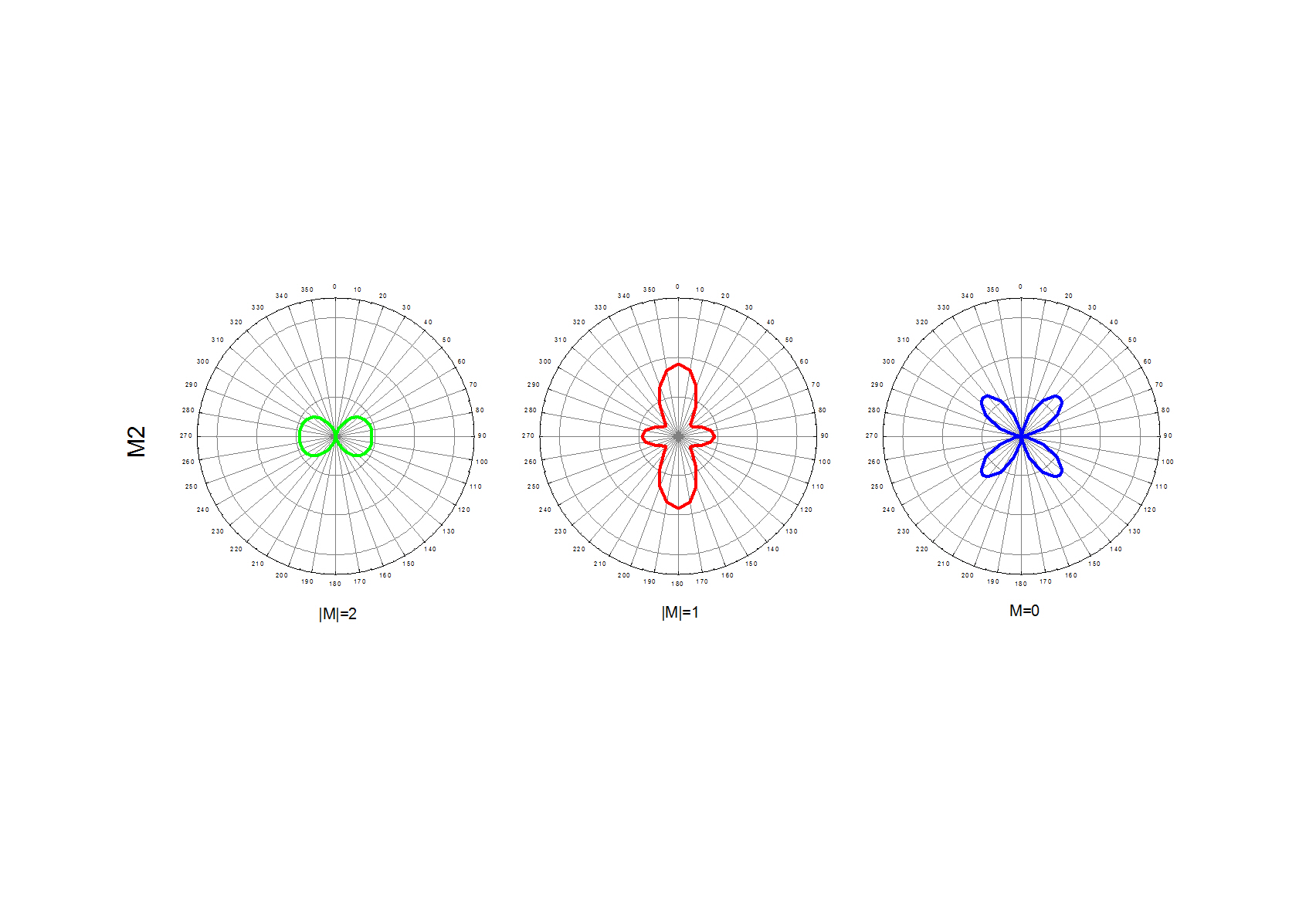}
\vspace{-1cm}
\includegraphics[scale=0.6]{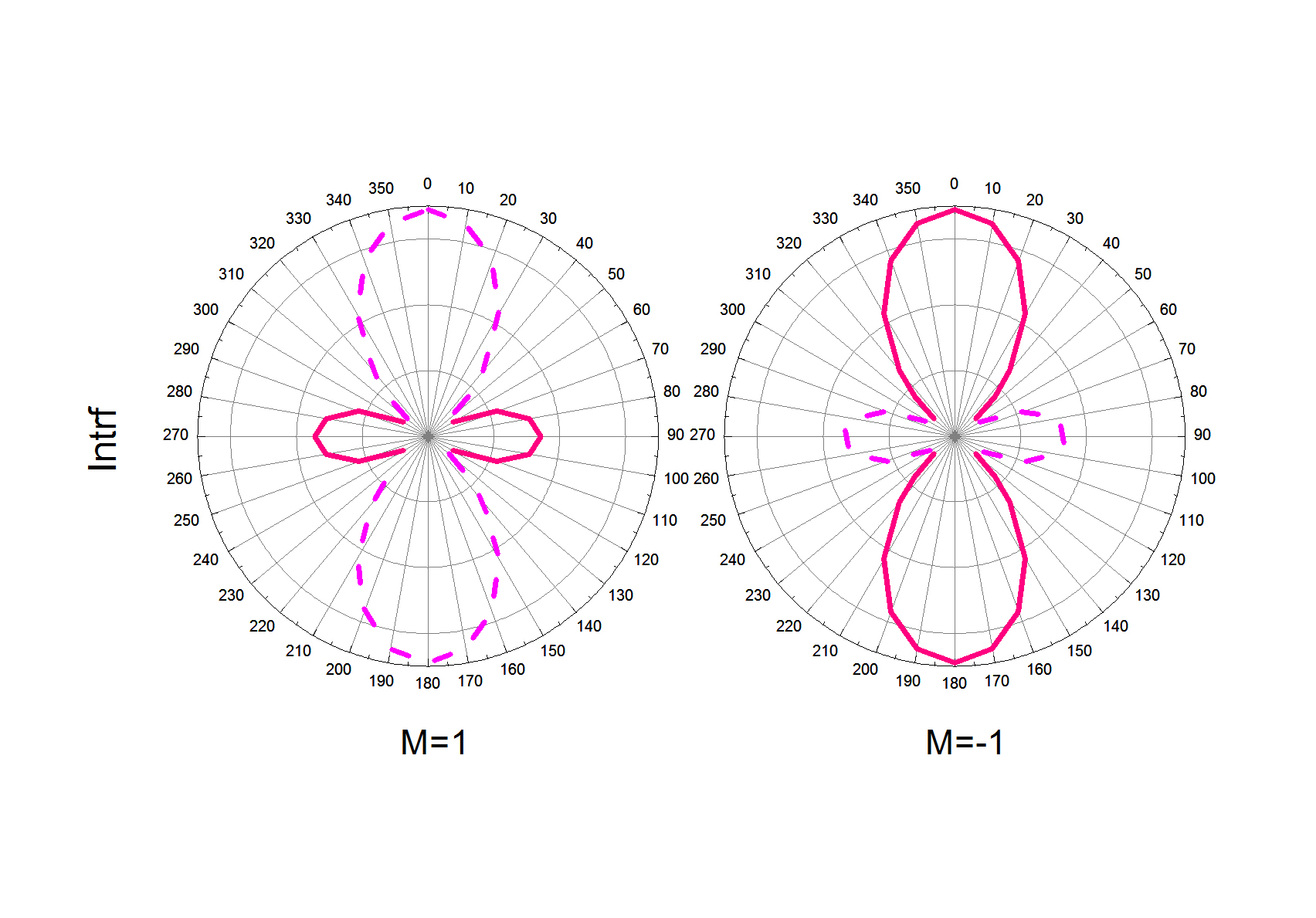}
\caption{\label{ADTR} The angular distributions of rates, without distinguishing the photon's polarization, for the magnetic-field induced E1 transition, M2 transition and the contribution from the interference (labelled with Intrf) between these two transitions in Ne-like Mg. 0$^{\circ}$ stands for the direction of the external magnetic field. The $M$ values are magnetic quantum numbers of the photon produced by the corresponding transition. The transition rates in atomic unit are magnified by a factor of $10^{17}$. The dash lines represent negative value.}
\end{figure}

Taking an example, we illustrated the angular distributions of the rates in Figure~\ref{ADTR} for the transitions in Ne-like Mg under the circumstance of $B=1$ T. The transition amplitudes $T_{if}$ were taken from Ref. \cite{Jonsson2013a, Li2013}. The rates in atomic unit were enlarged by a factor of $10^{17}$. Since the same scale was used in the figures, the relative magnitude for different terms in Eq. (\ref{ADTP}) can be seen. We found that the contributions from the interference term are larger than those from the magnetic-field induced E1 and M2 transitions, and thus essentially significant. In addition, the angular distributions of the interference term, unlike the E1 and M2 transitions, are dependent of the magnetic quantum number rather than its absolute value. For the present case, the interference term from the $M=1$ sublevel is positive along the direction of the external magnetic field but negative in the direction perpendicular to the field. For the other from $M=-1$ it has the opposite sign.   

\begin{figure}[!th]
\includegraphics[scale=1.0]{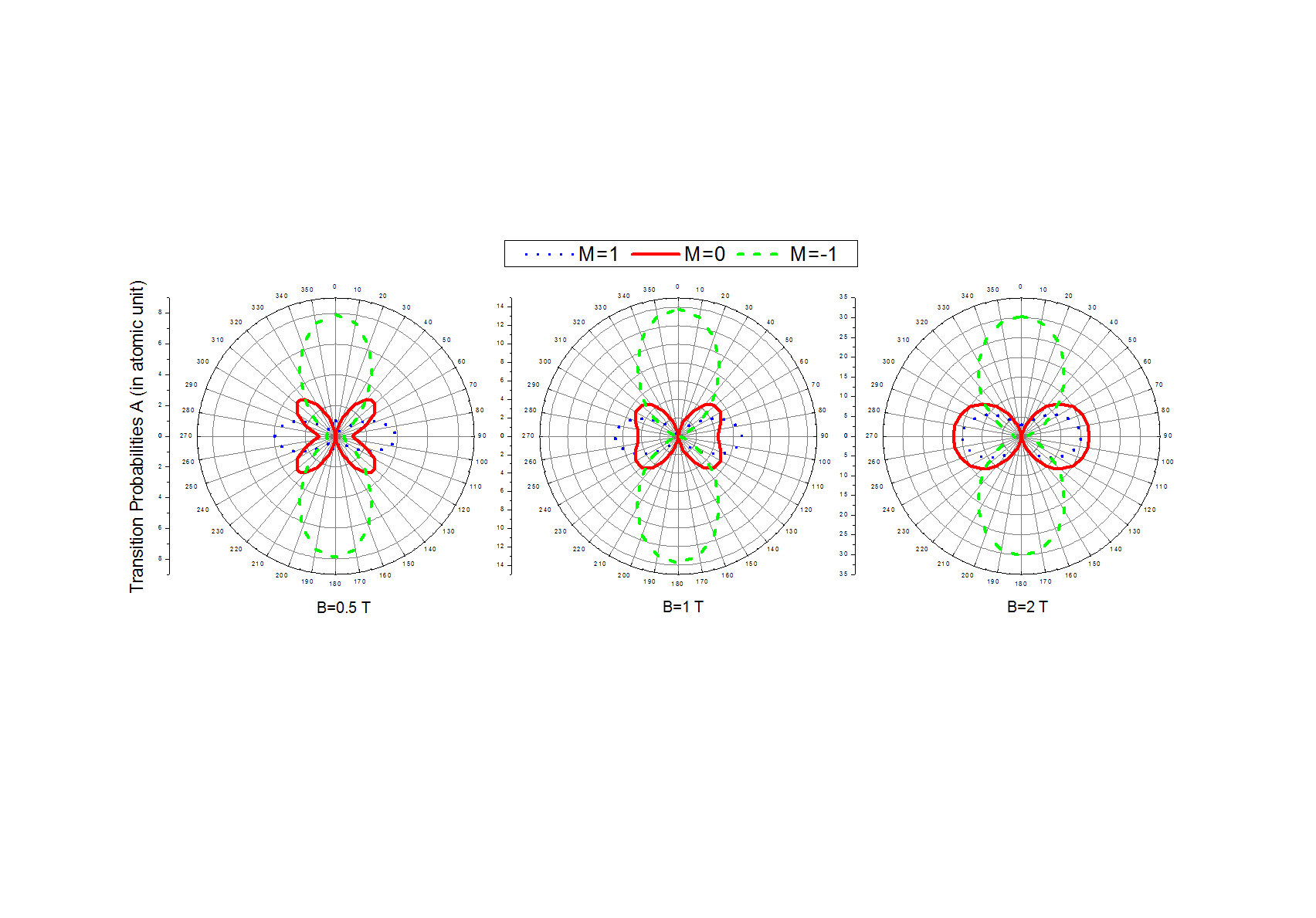}
\vspace{-3cm}
\caption{\label{ADTA} The angular distributions of the total probabilities, under the circumstance of $B=0.5, 1$ and $2$ T, for the transitions from $M=0, 1$ and $-1$ magnetic sublevels in $2p^53s~^3P_2$ state to the $2p^6~^1S_0$ ground state in Ne-like Mg.  $0^{\circ}$ stands for the direction of the external magnetic field. The transition rates are in atomic unit and magnified by a factor of $10^{17}$.}
\end{figure}

The angular distributions of the total transition rates were depicted in Figure~\ref{ADTA} for the $M=0, \pm 1$ three sublevels in $2p^53s~^3P_2$, because these three transitions are related to the magnetic field. As mentioned earlier, the magnetic-field induced E1 transition and the interference term are proportional to $B^2$ and $B$, respectively, which lead to the magnetic-field-dependent angular distributions. To show this effects, we also presented the angular distributions under the circumstance of $B=0.5, 1$, and $2$ T. As seen from Figure~\ref{ADTA}, the angular distributions of the emissions from $M=1$ and $M=-1$ are remarkably different by virtue of the contrary interference effect in these two transitions. Additionally, we notice that the pattern of the angular distribution for the photon with M=0 is strongly dependent on the strength of the external magnetic field. This results from the competition between the magnetic-field induced E1 and M2 transition and the fact that the MIT rate is proportional to the square of the magnetic field strength.      

Apart from the anisotropic angular distributions, the linear polarization of the radiation field is affected by the external magnetic field. In atomic physics, the linear polarization is specified by its degree~\cite{Surzhykov2003}
\begin{equation}
\label{LPD}
P(\theta) = \frac{I_{\parallel}(\theta) - I_{\perp}(\theta)}{I_{\parallel}(\theta) + I_{\perp}(\theta)}.
\end{equation} 
Here, $\theta$ denotes the angle of the photon's propagation direction ($\vec{k}$) with respect to the external magnetic field. $I_{\parallel}$ and $I_{\perp}$ refers to the intensities of light linearly polarized in the parallel and perpendicular direction, respectively,  in the plane that is orthogonal to the photon's propagation direction. For simplicity but no loss of generality, we calculated the linear polarization degree for the emission with $M=0$. According to Eq. (\ref{LPD}), the linear polarization degree can be expressed in this form
\begin{equation}
P(\theta) = \frac{\Delta - 5\cos^2\theta}{\Delta + 5\cos^2\theta},
\end{equation} 
where $\Delta$ stands for the ratio of rates between the magnetic-field induced E1 and M2 transitions and therefore is associated with the magnetic field strength. This also results in the dependence of the linear polarization degree $P$ on the magnetic field strength. To show this effect, we illustrated in Figure ~\ref{LPD} the linear polarization degree for Ne-like Mg under circumstance of $B=0.5, 1$ and $2$ T. It was found that the linear polarization degree is indeed sensitive to the magnetic field strength in this case.

\begin{figure}
\includegraphics[scale=1.0]{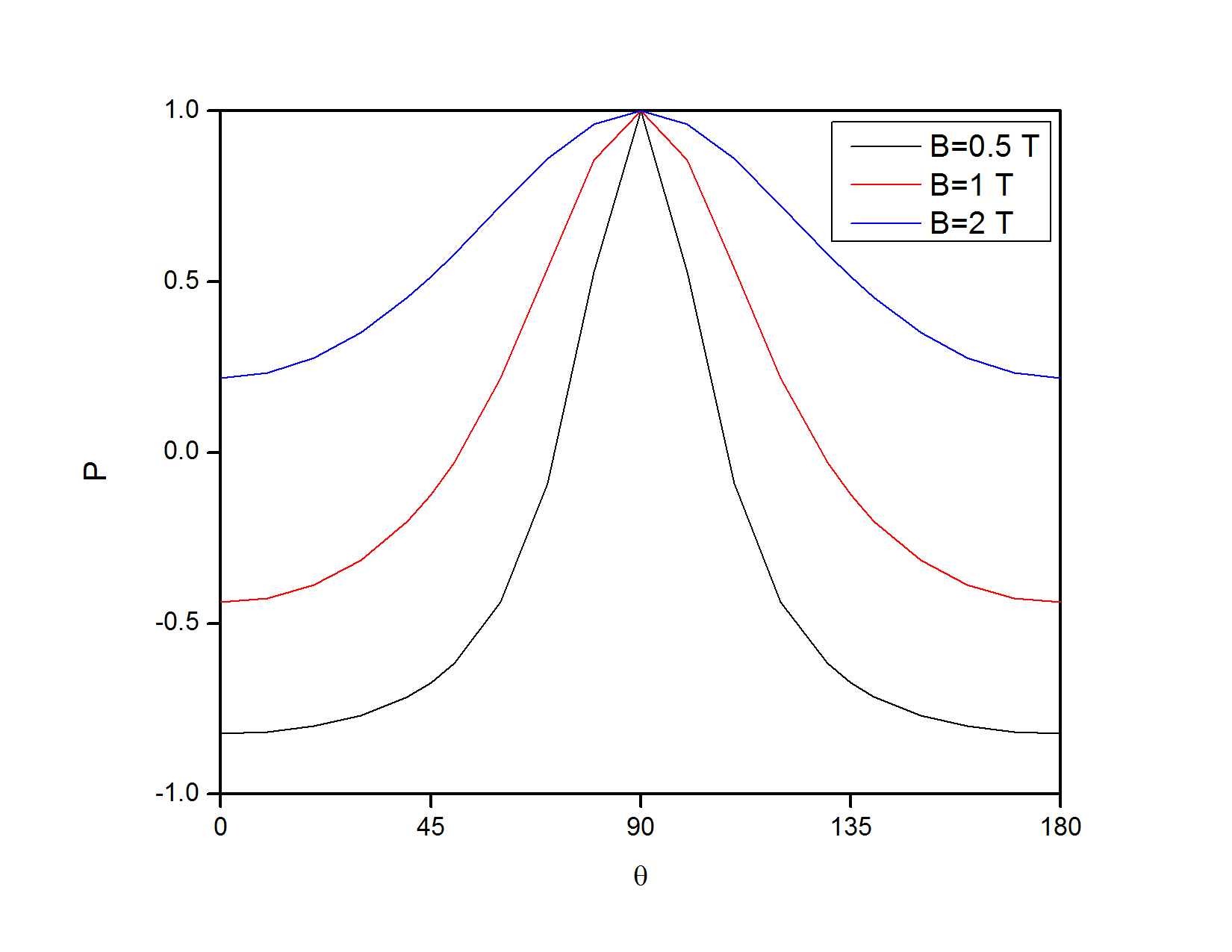}
\caption{\label{LPD} The linear polarization degrees of the emission with $M=0$ in the Ne-like Mg ion under circumstances of $B=0.5, 1$ and $2$ T.}
\end{figure}

In summary, we studied the effect of an external magnetic field on the angular distributions and the linear polarization of emissions from magnetic sublevels of the $2s^22p^53s~^3P^o_2$ state to the ground state in Ne-like ions. It was shown that these two physical quantities strongly depend on the magnetic field strength for the Ne-like Mg ions. Moreover, the effects can be seen for other Ne-like ions at the beginning or even in the middle of isoelectronic sequence where the magnetic-field induced E1 transition is comparable with the M2 transition. With respect to the magnetic-field-dependent angular distributions and linear polarizations, it may be considered as a tool to diagnostics of the magnetic field in plasmas.

\bibliography{MFAD}

\end{document}